\documentclass[epsf,prb,preprint,11pt,amsmath]{revtex4}
\usepackage{graphics}
\usepackage{graphicx}
\usepackage{dcolumn} 
\usepackage{bm}
\usepackage{epsfig}

\begin{document}
~
\vskip 10mm
\title{Magnetic Moment Collapse-Driven Mott Transition in MnO }

\vskip 10mm
\author{Jan Kune\v{s}$^{1,2}$}
\email{jan.kunes@physik.uni-augsburg.de}
\author{Alexey~V.~Lukoyanov$^3$}
\author{Vladimir~I.~Anisimov$^4$}
\author{ Richard~T.~Scalettar$^5$}
\author{ Warren~E.~Pickett$^5$}

\vskip 10mm
\affiliation{$^1$Theoretical Physics III, Center for Electronic Correlations and Magnetism, Institute of Physics,
University of Augsburg, Augsburg
86135, Germany}
\affiliation{$^2$Institute of Physics,
Academy of Sciences of the Czech Republic, Cukrovarnick\'a 10,
162 53 Praha 6, Czech Republic}
\affiliation{$^3$Ural State Technical University-UPI,
620002 Yekaterinburg, Russia}
\affiliation{$^4$Institute of Metal Physics, Russian Academy of
Sciences-Ural Division, 620041 Yekaterinburg GSP-170, Russia}
\affiliation{$^5$Department of Physics, University of California Davis,
         Davis, California 95616}


\date{\today}
\maketitle

{\bf The metal-insulator transition in correlated electron
systems, where electron states transform from itinerant to localized,
has been one of the central themes of condensed matter
physics for more than half a century.  The persistence of this 
question has been a consequence both of the intricacy of the
fundamental issues and the growing recognition of the complexities that
arise in real materials, even when strong repulsive interactions play the
primary role.  
The initial concept of Mott was based on the relative importance of kinetic
hopping (measured by the bandwidth) and on-site repulsion of electrons.
Real materials,
however, have many additional degrees of freedom that, as is recently  
attracting note,
give rise to a rich variety of scenarios for a ``Mott transition.''
Here we report results for the classic correlated insulator MnO which 
reproduce a simultaneous moment collapse, volume collapse, and metallization
transition near the observed pressure, and identify the mechanism as
collapse of the magnetic moment due to increase of crystal field splitting,
rather than to variation in the bandwidth.  
}

\vskip 3mm
We consider, as one of the simpler examples of the canonical Mott
insulators,\cite{mott,mott2}
the rocksalt structure transition metal monoxide (TMMO) manganese oxide
with half-filled $3d$ shell.  MnO
is, most certainly, a multiorbital multielectron
system with the accompanying
complexities of the tenfold degeneracy, but the half-filled $3d$ states
under ambient conditions lead to a spherical spin-only magnetic moment.  Applying
pressure to such a system leads to a number of changes, including
insulator-metal transition, orbital repopulation,
moment reduction, and volume collapse if a
first-order transition results.  These changes
may occur simultaneously, or sequentially over a range of volumes.\cite{RMP}  Any
of these may be accompanied by a structural phase transition, that is, a
change in crystal symmetry, but an isostructural volume collapse may
occur as well.  The $3d$ bandwidth of such a Mott insulator is very
susceptible to applied pressure, and has been thought to be
one of the main controlling
factors in the transition.

While MnO's half-filled shell seems to
give it a connection to well studied models, 
this aspect also makes it
atypical of transition metal monoxides, as shown by Saitoh {\it et al.} who
compiled\cite{saito95a} effective parameters for TMMOs from
spectroscopic information.  An effective intra-atomic Coulomb repulsion
energy $U_{eff}$ as defined by them is roughly twice as large as for
the other $3d$ monoxides, and this has been used to suggest that MnO may be the
most strongly correlated TMMO.
The complexity that should be expected can be grasped by listing the relevant
energy scales: $3d$ bandwidth $W$, Coulomb repulsion $U$,
intra-atomic $d-d$ exchange energy (Hund's rule) $J$, crystal field
splitting $\Delta_{cf} = \varepsilon_{e_g} - \varepsilon_{t_{2g}}$, and
charge transfer energy $\Delta_{ct} \equiv \varepsilon_{t_{2g}} -
\varepsilon_p$ [the difference in Mn $3d$ (we use $t_{2g}$) and O $2p$ site
energies]. All of these scales evolve as the
volume decreases, altering the various microscopic processes and
making the pressure-driven Mott transition a
challenging phenomenon to describe.

Early shock data\cite{noguchi}, and then Raman and optical studies,\cite{mita,mita2}
had identified a transformation in MnO in the neighborhood of 90-105 GPa.
Transport,\cite{patterson} magnetic,\cite{patterson} structural and 
spectroscopic,\cite{yoo,rueff} and reflectivity\cite{mita}
data all point to a first-order, insulator-metal Mott transition 
near 100 GPa with (reduced) volume 
($v=V/V_{\circ}$) collapse $v$=0.68$\rightarrow$0.63,
and moment collapse (from $\sim$5$\mu_B$ to 1$\mu_B$ or less\cite{yoo,rueff}).  
The structural data indicates
a B1$\rightarrow$B8 change just before the Mott transition, which thus occurs
within the B8 (NiAs) phase rather than the B1 (NaCl) phase.  Since the
local environment of the Mn ion remains the same, this structural change
is not expected to have much effect on the Mott transition in the
disordered phase.

\begin{figure}
\includegraphics[angle=270,width=\columnwidth,clip]{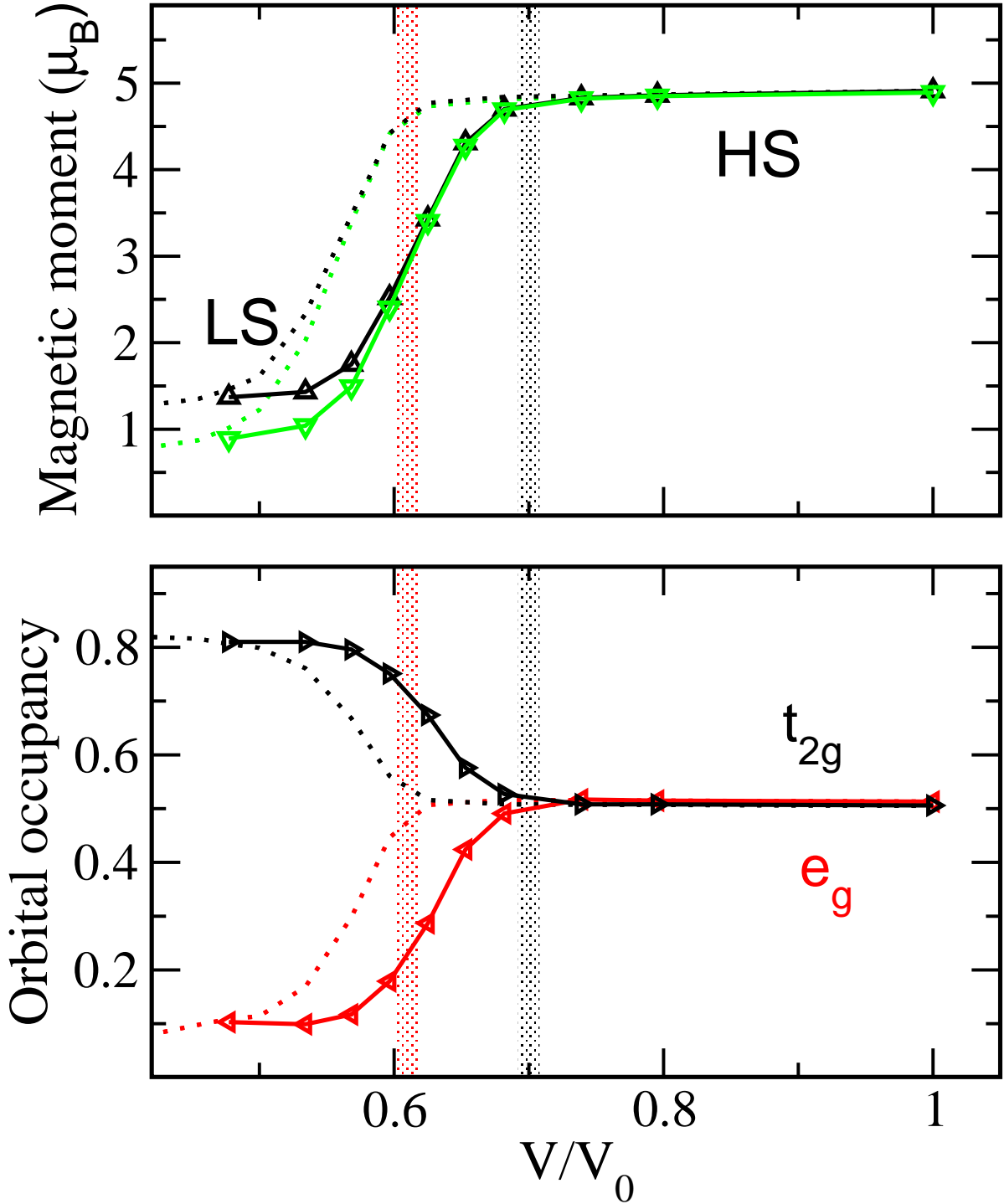}
\caption{\label{fig:collaps} This figure shows how the decrease in the
local moment correlates with the orbital occupations, which reveal the
spin state with relative volume.. 
The upper panel gives average instantaneous local moment $M_s$ (black) and
effective local moment $M_{eff}$ (green), and comparison to the Mn $3d$ orbital
occupancies (lower panel) resolved into $e_g$ (red) and $t_2g$ (black) components.
The solid lines represent the results obtained with the physical values
$U$=6.9 eV, $J$=0.86 eV; the dashed lines using the enhanced value $J$=1 eV
and constant $U/J$ ratio illustrates how the moment collapse is suppressed
to smaller volume if 
the spin-exchange coupling is increased.
Closing of the $t_{2g}$ and $e_g$ gaps 
is indicated by the black and red
vertical lines respectively, confirming a connection between metallization
and moment collapse.
The analogous closing of the gaps for the dotted line case ($J$=1 eV)
is shifted 
correspondingly (not shown
here).}
\end{figure}

Dynamical mean field theory (DMFT)\cite{metz,DMFT1,PT} as an approach for studying real materials
has been showing impressive 
successes.\cite{lich,heldCe,savkot,heldv2o3,georges,ldadmft,DMFT2,kun07,kuno7b}
The method that we have implemented and applied (see Methods section below)
moves the treatment significantly beyond the
methods used earlier for TMMOs, by including a full thermodynamic 
average of local dynamic
processes resulting from the strong interaction and all orbitals that
can be relevant.  
Cohen, Mazin, and Isaak calculated the energy and magnetic
moment using only local density approximation (LDA) based interactions.\cite{cohen} 
In LDA MnO metalizes at (much too) low pressure; within the metallic phase 
they obtained a moment and
volume collapse around 150 GPa.  Fang and collaborators addressed this difficulty
by using LDA only for the high pressure phase, and modeling the low pressure phase
with the correlated LDA+U method.\cite{fang}  
With two different functionals, however, it
is not possible to obtain the transition pressure.  Four correlated electronic structure 
methods \cite{deepa}, applied throughout the volume range of interest,
have probed the behavior of MnO under pressure; all obtained a high spin 
(HS, $S=\frac{5}{2}$) to low spin (LS,
$S=\frac{1}{2}$) moment collapse but their predictions differed considerably
in other respects, demonstrating that the specific treatment of correlation
effects is crucial.  The prediction of the LDA+U method,
which is regarded as the static, T=0
limit of the LDA+DMFT theory used here, is found to be affected by 
magnetic order,\cite{deepa2} and predicts a zero temperature moment collapse in an
insulator-insulator transition around 120 GPa (the pressure depends 
on the value of $J$),
with little difference between the B1 and B8 structure results.  
Thermodynamic fluctuations have not
been included in any previous study of MnO.  

\begin{figure}
\includegraphics[angle=0,width=\columnwidth,clip]{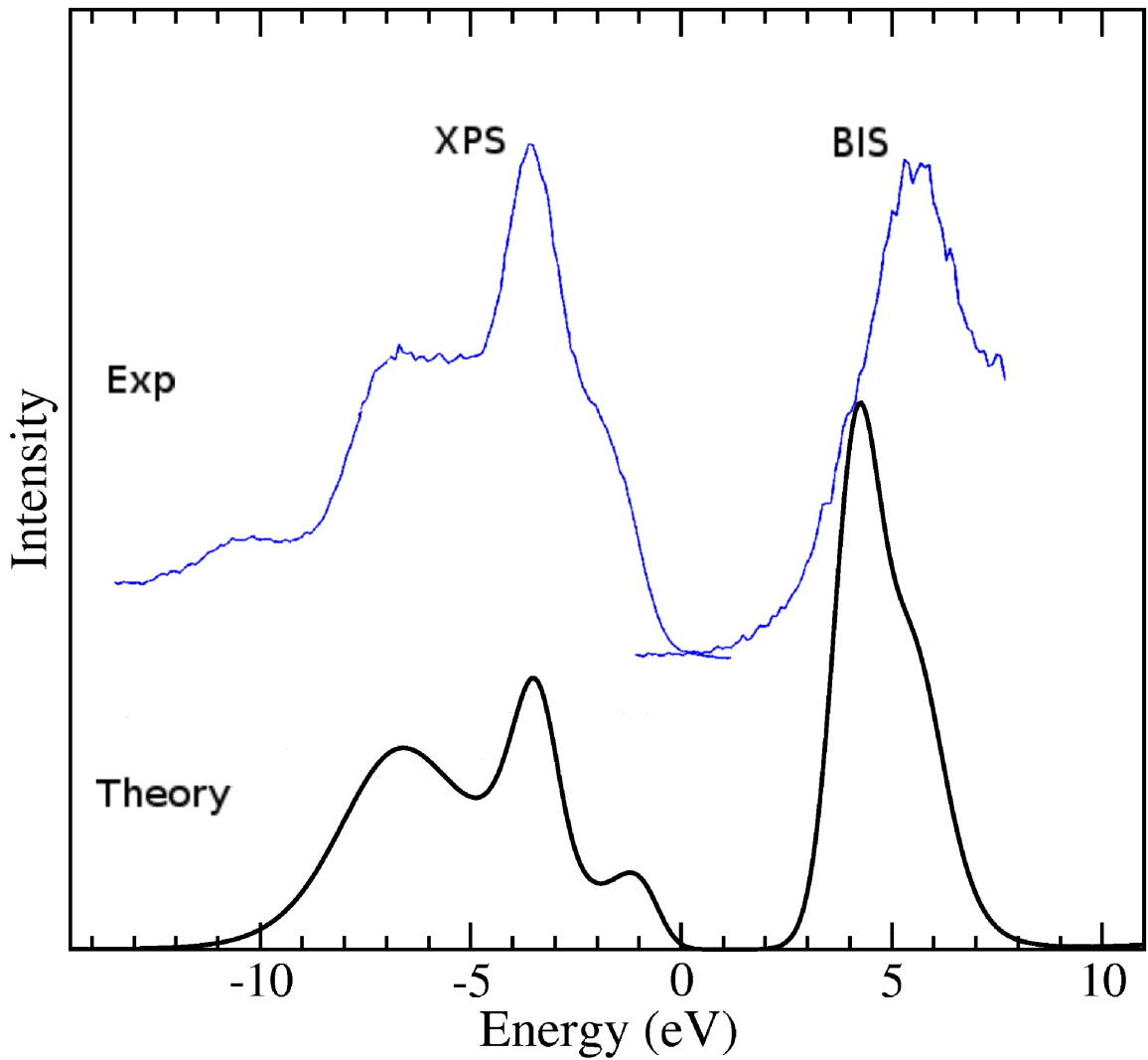}
\caption{\label{fig:xps} Ambient pressure x-ray photoemission spectroscopy (XPS) 
and bremsstrahlung isochromat spectroscopy (BIS) data of
van Elp {\it et al.}[\onlinecite{elp}] on both sides of the energy gap for MnO.
The  upper curve, offset for clarity, is to be compared with the 
present DMFT result (bottom curve).
While the separation of the main peaks is underestimated by $\sim 10$\%,
the overall agreement in positions of structure is excellent.
}
\end{figure}

\vskip 7mm \noindent
{\Large {\bf Magnetic Moment Collapse and Metallization}}\\ \vskip 3mm
Following most closely the approach developed and implemented by McMahan,
Held, and Scalettar\cite{mcm0,mcm1,mcm2} for pressure studies of elemental lanthanides, we have
addressed the pressure-driven collapse of the correlated insulating state, using
MnO as the prototype.
Fig. \ref{fig:collaps} illustrates
the evolution of the local magnetic moment and Mn $3d$ occupancies
with volume. We use two different measures of the local 
moment: (a) the mean instantaneous moment defined as an equal time
correlation function $M_s=\sqrt{\langle \hat{m}_z^2 \rangle}$ and
(b) effective local moment defined through the local spin susceptibility
$M_{eff}=\sqrt{T\chi_{loc}}$.  These two moments have similar T-independent values 
in materials with Curie-Weiss behavior.  
Under compression, the local moment and Mn $3d$ orbital occupancies
retain their ambient pressure HS values ($S=\frac{5}{2}$)
down to about $v$=0.68.
Further compression rapidly degrades the moment, which
is accompanied
by redistribution of electrons $e_g\rightarrow t_{2g}$ 
within the Mn $3d$ shell.
The local moments and orbital occupancies start to level off to the LS values
around $v=0.57$.
The reduction of $M_{eff}$ below $M_s$ in the LS state
indicates that the local moment screening (charge fluctuations) is enhanced
in comparison to the HS state.

\begin{figure}
\includegraphics[angle=270,width=\columnwidth,clip]{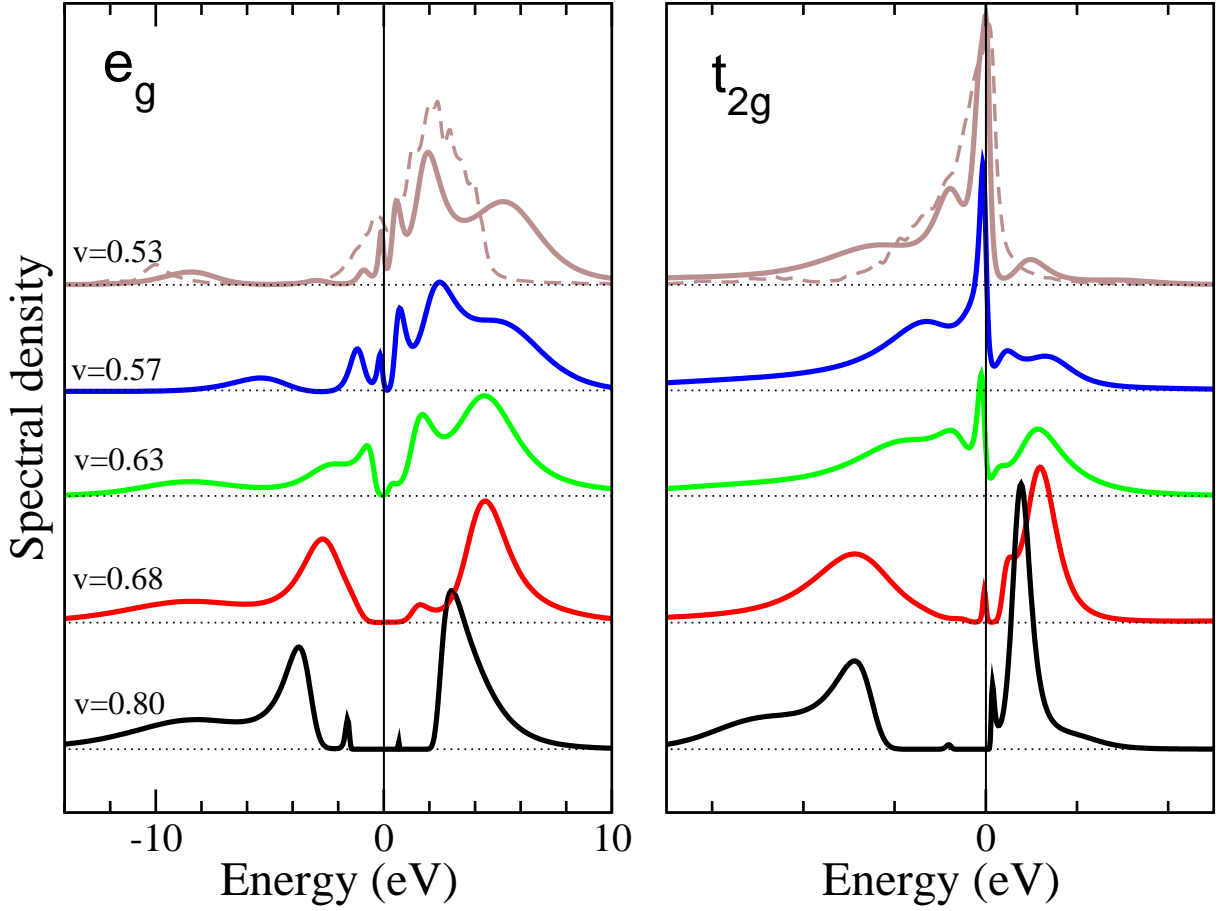}
\caption{\label{fig:spec}  View of the evolution of the Mn $3d$ spectral
densities under pressure (pressure increasing from bottom to top).
The single-particle
spectral functions are resolved into $e_g$ (left) and  $t_{2g}$ (right)
irreducible representations for varying relative volume.
Note the spectral weight shift under pressure: de-occupation 
of $e_g$ occurs as the increase in occupation of $t_{2g}$ proceeds 
(occupation $\equiv$ integrated weight over negative energies).
For the lowest volume we show the uncorrelated (LDA) spectra for comparison
(dotted lines).
Apparently the main spectral features at high pressure originate 
from the uncorrelated band structure with some many-body renormalization.
At even higher pressures the spectra remain qualitatively unchanged
with some reduction of the weight of the high energy shoulders.}
\end{figure}

Next we address the spectral properties, where the shortcomings of the LDA
spectrum have been clear for decades.  In Fig. \ref{fig:xps} we compare
the calculated total Mn $3d$ spectral function at ambient pressure with
the photoemission data of van Elp {\it et al.} \cite{elp}. 
Excellent agreement is obtained for the gap and for the peak positions.  
(We note that using the enhanced value of $J$=1 eV gives significantly poorer agreement.)
Having obtained a correct ambient pressure spectrum, we proceed in the
study of the Mott transition
by following the evolution with decreasing volume of the symmetry-resolved 
($t_{2g}, e_g$) spectral densities, presented in Fig. \ref{fig:spec}.
The onset of the moment collapse around $v=0.68$ is signaled by, and
associated with, closing
of the gap in the $t_{2g}$ channel, while the $e_g$ gap
is still visible at $v=0.63$. 
This orbital selectivity\cite{orbselect1,orbselect2} in metalization cannot be an exact 
property since both $e_g$ and $t_{2g}$ bands hybridize with the same
O $2p$ bands throughout the Brillouin zone; however, the smallness of $t_{2g}$--$2p$
mixing allows the orbital selectivity to be remarkably pronounced.
As the $t_{2g}$ gap closes, a quasiparticle peak appears at the chemical potential
(E=0) as has been seen in simple models.\cite{DMFT1}
Once in the LS state, the spectral functions bear strong resemblance
to the parent LDA bands.  In particular, the LDA ($U=J=0$) $t_{2g}$ spectrum
contains a sharp peak just at/below the chemical potential, so it is not
certain how much of the peak arising at the transition is due to the
many-body nature of the system.

\begin{figure}
\includegraphics[angle=270,width=\columnwidth,clip]{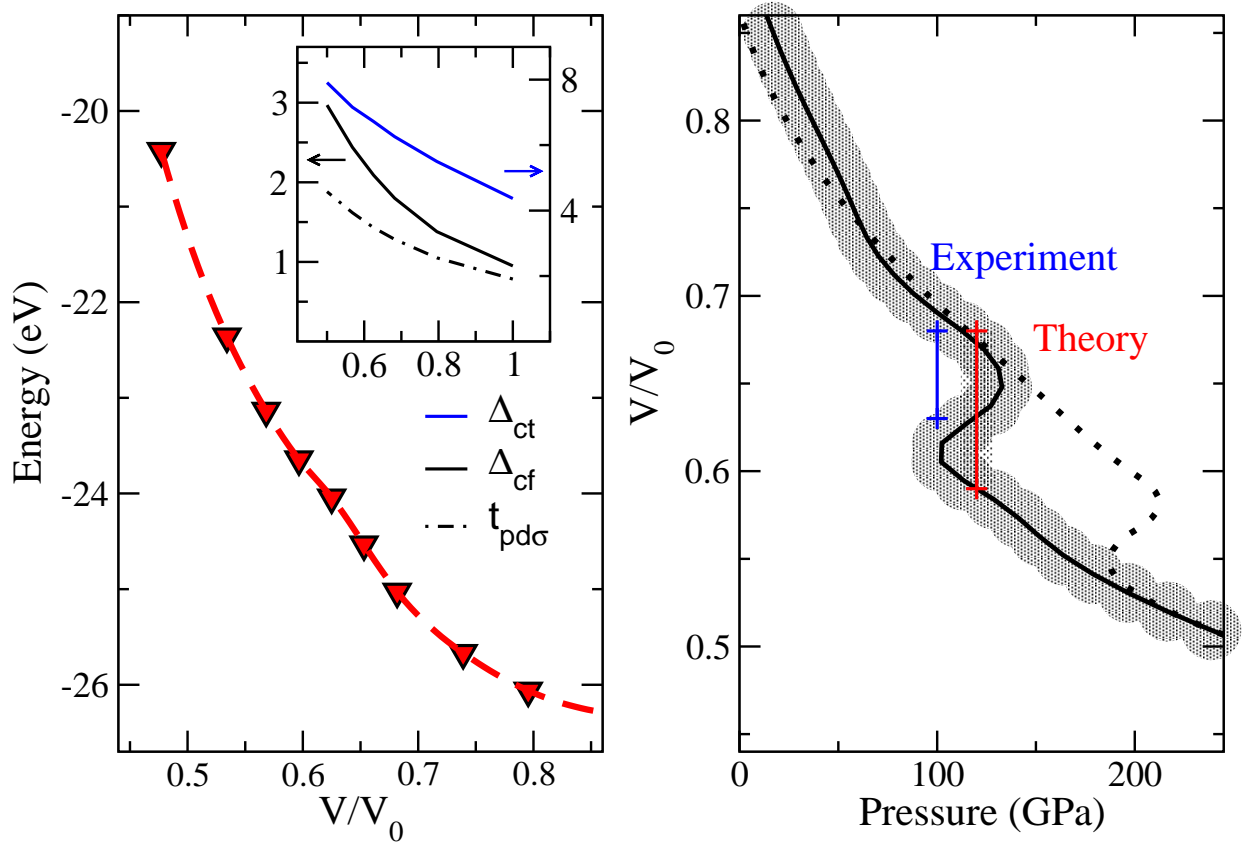}
\caption{\label{fig:ene} The equation of state that quantifies the volume collapse
transition is presented in two representations.  In the left panel is the
internal energy versus volume (dashed line represents a spline interpolation), and
the right panel shows the resulting volume versus pressure curve
(shading indicates estimated uncertainty), obtained
as a derivative of the spline interpolation of E(V).
The red bar on the $V(P)$ curve lying at the theoretical transition 
pressure $P^{th}_c$
= $120\pm15$ GPa determines the volume collapse $v=0.68\rightarrow 0.59$.
The dotted curve represents $V(P)$ for the enhanced value of exchange $J$=1 eV,
showing the shift of the Mott transition to higher pressure with larger $J$.
Width of the shaded red bar indicates the uncertainty of $P^{th}_c$ due to 
stochastic nature of QMC.
The inset in the left panel shows the evolution of selected tight-binding
parameters (units of eV); note specifically the factor of three 
increase in $\Delta_{cf}$ .
}
\end{figure}

\vskip 7mm \noindent
{\Large {\bf Mechanism of the Mott Transition in MnO}}\\ \vskip 3mm
We now address a fundamental point of this work, namely the
connection between moment collapse and metal-insulator transition,
by observation of the impact of pressure on the effective Hamiltonian.
Since $U$ and $J$ do not change, the pressure enters the calculation
only through the quadratic (one-electron) part of the effective Hamiltonian.
Reducing the role of pressure down to fundamentals one ends up with
two effects: (i) broadening of the $3d$ bands and (ii) increase of
the crystal-field splitting $\Delta_{cf}$. (We define $\Delta_{cf}$ in terms
of the site energies of the $e_g$ and $t_{2g}$ Wannier functions; the
$e_g$--$t_{2g}$ band splitting is substantially larger due to ligand
field effects.)  The evolution of the leading
band structure quantities, which are the nearest-neighbor hopping amplitude
$t_{pd\sigma}$, $\Delta_{cf}$, and $\Delta_{ct}$,
are shown in the inset of Fig. \ref{fig:ene}.
Since the $3d$ bandwidth arises mainly through Mn $3d$ -- O $2p$
hybridization ($W\propto t_{pd}^2/\Delta_{ct})$ the increase of 
$t_{pd}$ hopping with pressure is to some
extent compensated by the overall lowering of the $p$ bands (increase
in $\Delta_{ct})$.

So far we have demonstrated a connection between the
moment collapse and metal-insulator transition (MIT), 
yet the chicken-and-egg question -- which property drives? which property follows?
-- is not yet answered.
To this end we have performed an additional calculation at $v=0.8$
(well within the insulating HS state) without {\it any} intra-atomic exchange ($J$=0).
In spite of the large $U$ and same $U/W$ ratio, a LS solution is obtained,
which is metallic although strongly renormalized.
This result clearly shows that the MIT is {\it driven by the collapse of
the moment}, which cannot withstand the increase of $\Delta_{cf}$.
The transition is characterized as evolving from five half-filled
bands $t_{2g}+e_g$ (HS) to three $t_{2g}$ bands with one hole per site (LS).
The interaction energy cost of moving an electron from site to site
is determined by  
$U_{eff}=d^{n+1}+d^{n-1}-2d^n$. Using the atomic configurations corresponding
to HS and LS states one arrives at an {\it effective} repulsion\cite{ani91}
$U_{eff}^{HS} = U+4J =$ 
10.3 eV and $U_{eff}^{LS} = U-J =$ 5.9 eV
respectively, indicating much stronger inhibition of the electron 
propagation in HS state. Moreover, the Mn $3d(e_g)$ -- O $2p$ hybridization 
provides an additional screening channel for the effective 
$t_{2g}-t_{2g}$ interaction
in the LS state.
Indeed, a calculation performed in the LS state with O $2p$ states 
integrated out (keeping the $3d$ bandwidth unchanged) before solving the interacting 
problem leads to more pronounced high energy shoulders as compared
to the solution with O-$p$ states fully included; an indication of stronger
local correlations. 

\begin{figure}
\includegraphics[angle=0,width=\columnwidth,clip]{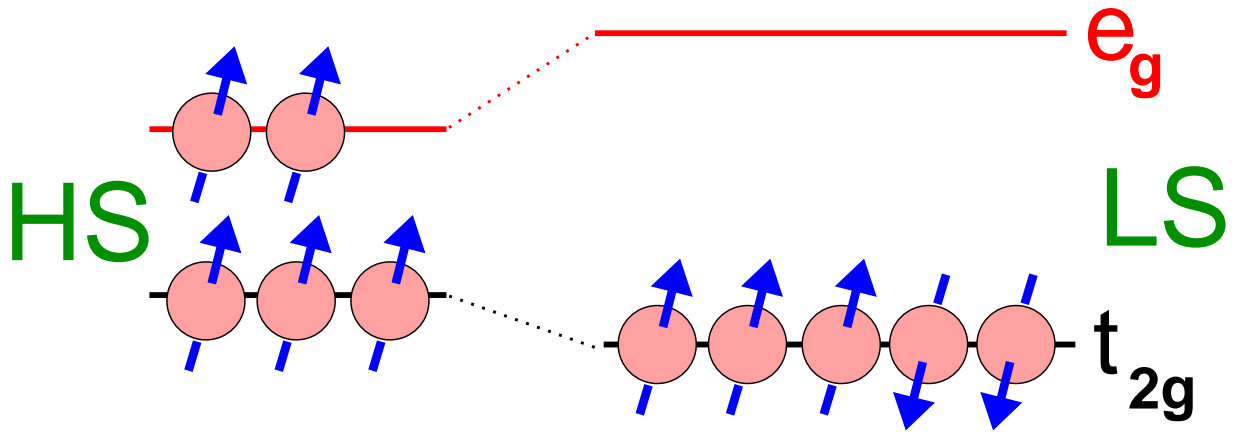}
\caption{\label{fig:cartoon} Schematic energy diagrams of the spin states
at both ambient pressure and at high pressure in the collapsed phase.
Occupations of the Mn $3d$ levels are pictured for both high spin
(left) and low spin (right).  In the HS
state two spin-up electrons occupy $e_g$ orbitals at the cost of 2$\Delta_{cf}$
in energy, but the spin-exchange energy gain is -10$J$ (5$\times$4/2 = 10
pairs of parallel spin electrons).  In the LS state, the crystal field energy
cost has become too great, and although the spin-exchange energy is less
[-4$J$ from 3$\times$2/2 (up) + 1 (down) = 4 pairs] there is a net energy
gain.  The LDA energy difference is also a factor.
}
\end{figure}
These results establish that the transition is controlled
by competition between the crystal-field spitting $\Delta_{cf}$ (favoring the
LS state) and 
the intra-atomic exchange coupling $J$ (favoring the HS state).  Although both
energy scales are important for the outcome of the calculations, 
only the former ($\Delta_{cf}$) is sensitive to an applied pressure.
The importance of the value of $J$ was also found in LDA+U
studies of the Mott transition.\cite{deepa2}
Recently Werner and Millis \cite{0704.0057} studied a two band model
with competing intra-atomic exchange and crystal-field splitting.
In the parameter range relevant for the present study
they found three different phases realized in the
following order with increasing crystal-field splitting:
(i) spin-polarized Mott insulator,
(ii) metal with large orbital and spin fluctuations and
(iii) orbitally polarized insulator.
Moreover they found an orbitally selective closing of the gap upon doping 
in the vicinity to the (i)-(ii) phase boundary.
The correspondence between their spin-polarized insulator phase (i) and the HS state of MnO is 
evident. The transition region in MnO and phase (ii) are both  characterized by metallization
and strong orbital fluctuations as well as the orbitally selective gap behavior.
Also the LS state of MnO and the phase (iii) of the model exhibit similarity, the orbital
polarization. The insulating character of their phase (iii) is dictated by band-filling
and does not contradict the above analogy.

Like almost all previous studies using LDA+DMFT, we have included only the density-density
terms of the Coulomb repulsion.  Although they are not expected to influence a first-order
volume collapse (see the next section) especially above 1000 K, it is gratifying to obtain
some confirmation.  Werner and Millis used the full Coulomb interaction in their study, 
and the similarity of the behavior of their model to what is found here for MnO provides
some verification of the unimportance of the neglected terms.

\newpage
\vskip 7mm \noindent
{\Large {\bf Equation of State and the Critical Pressure}}\\ \vskip 3mm
To compare to high pressure
experiments, knowledge of phase stability is needed, which can be
obtained from free energy vs volume (equation of state EOS). 
The theoretical justification for applying  DMFT using the underlying LDA  
description relies on a well-defined thermodynamic grand canonical potential 
functional, for which 
specific realizations
have been suggested.\cite{heldCe,mcm0,SYS}
Since it is difficult to extract
the entropic term\cite{mcm1} in the free energy 
we restrict ourselves to evaluation of the internal energy;
in any case the variation of the entropy term is very small on the energy scale 
of several eV involved in the changes of total energy. 
We use the internal energy scheme of 
McMahan {\it et al.}\cite{heldCe,mcm0,mcm1,mcm2} that is similar to that of Savrasov and 
Kotliar\cite{SYS} corresponding to the expression
 $ E(V,T)=E_{LDA}(V) + [E_{DMFT}(V,T) - E_{MF}(V)],$
where $E_{LDA}$ is the all-electron (unpolarized) LDA energy, 
$E_{DMFT}$ is the internal energy
corresponding to the self-consistent (dynamic) DMFT solution for the
effective Hamiltonian and $E_{MF}$ is the static mean-field
internal energy. The EOS 
curve is shown in Fig. \ref{fig:ene}. The main feature 
is the deviation from convexity in the transition region, which
leads to a calculated volume collapse $v^{th}=0.68\rightarrow 0.59$ 
at $P^{th}_c = 120$ GPa.
The metallization and moment collapse obtained here are not far from
the high pressure data,\cite{patterson,yoo,rueff} with the transition volume
(pressure) being somewhat smaller (larger) than the experimental values
$v^{exp}_c=0.68\rightarrow 0.63$, $P^{exp}_c$ = 100 GPa. 
The state just above the collapse is a HS insulator, with the $t_{2g}$ gap
about to close.  The collapsed state is LS, the $e_g$ gap having just closed
making it metallic in both subshells.

\vskip 7mm \noindent
{\Large {\bf Summary and Outlook}}\\ \vskip 3mm
These results demonstrate that the underlying LDA band structure, 
buttressed by on-site
interactions ($U$, $J$) treated within the dynamical DMFT ansatz, 
provide a realistic description of the Mott transition
in MnO without input from experiment.  
This study finally allows a 
determination of the mechanism of
the transition, which could not be uncovered by experiment alone:
the magnetic moment collapse, volume collapse, and metal-insulator transitions
occur simultaneously, but it is the increasing crystal field splitting 
(encroachment of the O$^{2-}$ ion on the internal structure of the Mn ion) 
and not the increasing bandwidth
that tips the balance. 

The current results illustrate
success of the LDA+DMFT approach in describing a pressure-driven Mott transition in a strongly
correlated insulator, joining the growing number of successes of this approach in other strongly
correlated real materials.  The Kondo volume collapse transition in Ce\cite{heldCe,mcm1}
and other elemental lanthanides,\cite{mcm2} and the realistic modeling of parts 
of the complex phase diagram\cite{savkot} and multiplet effects\cite{shick}
in Pu reflect the progress in correlated metals, with low temperature properties
(heavy fermion characteristics) remaining an imposing challenge.  Impressive progress has also
been demonstrated in the description of structurally-driven\cite{heldv2o3,pavarini}
and doping-driven\cite{craco} metal-insulator transitions in 
transition metal oxides.  Excitation spectra\cite{kun07,kun07b} of the charge-transfer 
compound NiO at ambient pressure, where
O $2p$ states are tangled with the $3d$ states, have shown excellent agreement with experiment. 
These results on MnO bring an additional class of materials
into the list of strongly correlated systems whose behavior is becoming understood
due to recent theoretical developments.

\vskip 14mm \noindent
{\Large {\bf Theoretical Approach and Numerical Methods}}\\
\vskip 3mm \noindent
{\it Single Particle Hamiltonian and Interaction Term}\\
The LDA+DMFT computational scheme\cite{ldadmft} in its present implementation,
applied previously to NiO,\cite{kun07,kun07b}
proceeds in two steps: (i) construction of an effective multi-band
Hubbard Hamiltonian $H$ via Wannier transformation from a converged
(unpolarized, metallic) LDA solution corrected for double-counting of the  on-site
interaction,
and (ii) self-consistent solution of the DMFT equations \cite{DMFT1,DMFT2} using the quantum
Monte-Carlo impurity solver.\cite{solver}
\begin{eqnarray}
\label{eq:ham}
H&=&\sum_{\mathbf{k},\sigma,\alpha,\beta}h_{\mathbf{k},\alpha\beta}^{dd}d_{\mathbf{k}\alpha\sigma}^{\dagger}
d_{\mathbf{k}\beta\sigma}+
 \sum_{\mathbf{k},\sigma,\gamma,\delta}h_{\mathbf{k},\gamma\delta}^{pp}p_{\mathbf{k}\gamma\sigma}^{\dagger}
p_{\mathbf{k}\delta\sigma}+
 \sum_{\mathbf{k},\sigma,\alpha,\gamma}h_{\mathbf{k},\alpha\gamma}^{dp}d_{\mathbf{k}\alpha\sigma}^{\dagger}
p_{\mathbf{k}\gamma\sigma}+
 \sum_{\mathbf{k},\sigma,\gamma,\alpha}h_{\mathbf{k},\gamma\alpha}^{pd}p_{\mathbf{k}\gamma\sigma}^{\dagger}
d_{\mathbf{k}\alpha\sigma}   \nonumber  \\
 & &   +\sum_{i,\sigma,\sigma',\alpha,\beta}
        {\cal U}_{\alpha\beta}^{\sigma\sigma'}n^d_{i\alpha\sigma}n^d_{i\beta\sigma'},\nonumber
\end{eqnarray}
where $d_{\mathbf{k}\alpha\sigma}$ ($p_{\mathbf{k}\gamma\sigma}$)
is the Fourier transform of the operator $d_{i\alpha\sigma}$ ($p_{i\gamma\sigma}$), 
which annihilates the $d$ ($p$) electron
with orbital and spin indices $\alpha\sigma$ ($\gamma\sigma$) in the $i$th unit cell,
and $n^d_{i\alpha\sigma}$ is the corresponding $d$ occupation number operator.

The single particle part of the Hamiltonian was obtained by a Wannier function projection method \cite{wf},
which amounts to a unitary transformation in the Hilbert space containing Mn $3d$, O $2p$ bands and the next lowest
empty (polarization) conduction band. The site energy of the Mn $3d$ orbitals was corrected for double counting of the
$d-d$ interaction by subtracting from the LDA site energy $\varepsilon_d$ a Hartree-like term
giving $\varepsilon_d^{\circ} = \varepsilon_d-
(N-1)\bar{U}n_{LDA}$, where $N=10$ is the total
number of orbitals per Mn site, $\bar{U}$ is the average Coulomb repulsion and $n_{LDA}$ is the average
occupancy per $d$-orbital.  Since the $p-d$ band separation $\Delta_{ct}$ in MnO, which is
to some extent influenced by the choice of the double-counting term, is rather large in the transition
region, small variation of $\Delta_{ct}$ will not change the results.

\vskip 3mm \noindent
{\it The Coulomb Interaction Matrix}\\
The on-site Coulomb interaction ${\cal U}_{\alpha\beta}^{\sigma\sigma'}(U,J)$ within the Mn $3d$ shell,
restricted to
density-density terms only, was expressed as usual\cite{sasha} in terms of the direct ($U$) and
exchange ($J$) interaction strengths related
to the Slater integrals $F_0, F_2, F_4$. The numerical values\cite{ani91} of $U$=6.9 eV
and $J$=0.86 eV were obtained by the constrained LDA
method\cite{constrained}.  Since they exhibit
only a small pressure dependence, these values were used for all volumes.
We used L=100 imaginary time points in the Monte-Carlo simulation, in which
the standard single-field-flip moves were augmented by special global
moves that played a
crucial role in ensuring ergodic sampling in the transition region.
To obtain an indication of the robustness of our results we perform, in
parallel with these {\it ab initio} interaction strengths, calculations with
an enhanced (by 15\%) value of $J$=1 eV (and fixed $U/J$ ratio).
All the presented results were obtained at the temperature T=1160 K, in
the rocksalt structure.

\vskip 3mm \noindent
{\it Monte Carlo Procedure; Introduction of Global Moves}\\
The DMFT equations were solved numerically on a Matsubara contour (using asymptotic
expansions for frequencies $\omega_n>500$ eV), and the k-space integrals were performed by summation over 3375 k-points in the
first Brillouin zone. The chemical potential was adjusted in each DMFT iteration to guarantee the total
electron count of $11\pm 10^{-6}$.
The impurity problem
was solved using the Hirsch-Fye QMC algorithm\cite{solver} modified for multiple orbitals.
The on-site interaction was decoupled using
a single binary Hubbard-Stratonovich auxiliary field $S_{\alpha\beta}(l)$ for each
pair of orbitals
$\alpha\beta$ and each of $L$ imaginary time slices (45 auxiliary fields for each time slice).

The key innovation in this application
to MnO in the transition regime was introduction of global Monte-Carlo moves 
in addition to the usual single-flips of the auxiliary fields.
These moves allow
for fluctuations between HS- and LS-like configurations, which are otherwise practically unreachable with the
standard single-auxiliary-field-flip moves.
The purpose of global moves is
to mimic transferring electrons between orbitals. In general there is no
straightforward relationship between a given configuration of auxiliary fields,
described by a binary $L$-vector $S_{\alpha\beta}$ and the occupancy of
orbitals. However, in the case of two atomic orbitals the probability
distribution is peaked around auxiliary field configurations corresponding to a
particular orbital occupancy, and flipping all fields corresponds to
swapping occupancies of the two orbitals \cite{sca}. This can be
generalized to multiple orbitals as follows.
To swap occupancies of orbitals $\alpha$ and $\beta$ one has to:
(1) flip fields in $S_{\alpha\beta}$, (2) for all remaining fields coupled
to orbitals $\alpha$ or $\beta$ swap the configurations
$S_{\gamma\alpha}\leftrightarrow S_{\gamma\beta}$.
Since the decoupling is anti-symmetric with respect to the ordering of
orbitals, auxiliary fields must be
flipped in step (2) whenever the order of orbitals changes between
$S_{\gamma\alpha}$ and
$S_{\gamma\beta}$.

Testing several types of the above moves we found that
only simultaneous moves of two electrons between $t_{2g}$ and $e_g$ orbitals
of opposite spin
(i.e. moves intuitively expected in LS$\leftrightarrow$HS fluctuations)
have appreciable acceptance.
The acceptance rate of the global moves was found to be
large only in the transition regime, which had been characterized by unusually slow convergence
of the DMFT cycle. We checked for the possibility of multiple solutions, but found none at the
temperature of these simulations. The numerical value
of the total energy, limited by the stochastic error of the $E_{DMFT}$ term, was converged to the
accuracy of 0.06 eV in the transition regime and 0.02 eV anywhere else.
The spectral densities were calculated by the maximum entropy analytic continuation technique\cite{mem}
applied to the imaginary-time Green functions from $4\times10^7-6\times10^7$ QMC-simulation sweeps collected
into 2000-20000 bins.

\newpage
\vskip 7mm \noindent
{\Large {\bf References}}\\

\vskip 7mm \noindent
{\Large {\bf Acknowledgments}}\\ \vskip 3mm
 J.K. gratefully acknowledges the Research
Fellowship of the Alexander von Humboldt Foundation.  We acknowledge
numerous discussions with D. Vollhardt and A. K. McMahan, and
useful interaction with K.-W. Lee during the latter stages of this work.
This work was supported by SFB 484 of the Deutsche Forschungsgemeinschaft
(J.K.), by the Russian Foundation for Basic Research
under the grants RFFI-06-02-81017, RFFI-07-02-00041 (V.I.A. and A.V.L.)
and the Dynasty Foundation (A.V.L.),
by DOE grant No. DE-FG02-04ER46111, and by DOE Strategic Science Academic Alliance
grant No. DE-FG01-06NA26204.
This research was also encouraged and supported by the U.S. Department of Energy's 
Computational Materials Science Network (J.K., R.T.S., and W.E.P.).

\vskip 7mm \noindent
{\Large {\bf Competing financial interests}}\\ \vskip3mm
The authors declare no competing financial interests.

\end{document}